# MovieMat: Context-aware Movie Recommendation with Matrix Factorization by Matrix Fitting

Hao Wang
*Ratidar.com*
Beijing, China
haow85@live.com

*Abstract*—Movie Recommender System is widely applied in commercial environments such as NetFlix and Tubi. Classic recommender models utilize technologies such as collaborative filtering, learning to rank, matrix factorization and deep learning models to achieve lower marketing expenses and higher revenues. However, audience of movies have different ratings of the same movie in different contexts. Important movie watching contexts include audience mood, location, weather, etc. Tobe able to take advantage of contextual information is of great benefit to recommender builders. However, popular techniques such as tensor factorization consumes an impractical amount of storage, which greatly reduces its feasibility in real world environment. In this paper, we take advantage of the MatMat framework, which factorizes matrices by matrix fitting to build a context-aware movie recommender system that is superior to classic matrix factorization and comparable in the fairness metric.

*Keywords—MovieMat, MatMat, recommender system, movie recommender system, matrix factorization*

## I. Introduction

Movie recommender systems is one of the mostlucrative business in the internet industry. Video companies such as YouTube, NetFlix, Tubi, among many other firms have been relying on movie recommendation as a core technical asset to boost company's revenues. One of the earliest classic movie recommender models is from YouTube [1], which takes advantage of random walk of user neighbors to recommend related videos. The company's technologies have undergone tremendous advancement along the years. Now, YouTube build their video recommendation with highly complex deep learning architectures [2] .

Movie recommendation is usually in short of contextual information. For many years, video companies rely on user click behavior and payment log to recommend movies. However, it is well known that contextual information is highly helpful in many recommendation scenarios, as movie watcher selects movies depending on his mood, location, weather and many other factors. Methods such as tensor factorization are frequently mentioned in literature to solve context-aware movie recommendation problem. Unfortunately, tensor factorization consumes an unbearable amount of storage that is not feasible in modern day commercial environment.

Tensor factorization by definition is a mapping from $R^n$ to $R^{n-1}$ , due to the discrepancy of dimensions between input and output, a tensor in $4^{th}$ or $5^{th}$ dimensions in input causes a large number of items to be stored. Take MovieLens Small Dataset as example, the dataset contians 610 users and 9724 movies. A $4^{th}$ degree tensor could be 610*9724*610*9724*3*4 bytes, which leads to 384 TB in data storage. In general, commercial environments have much larger datasets, which makes it impossible to store data input in any commercial cluster on this planet.

In 2021, Wang [3] proposed a new framework inspired by tensor factorization called MatMat. The framework replaces the user rating values by matrices to be fitted by products of user feature matrix and item feature matrix. The matrices to be fitted can incorporate contextual information or multi-task learning goals. The algorithmic framework is highly effective in space complexity, and serves as an ideal replacement for tensor factorization. In the publication, the author proposes a special case of MatMat using user popularity ranks and item popularity ranks as matrix side information and proves that the new framework outperforms the classic matrix factorization in both accuracy and fairness metrics.

In this paper, we propose a new context-aware movie recommendation model called MovieMat and its variant MovieMat+ based on the MatMat framework. We incorporate the location and audience mood into the matrices to be fitted and proves in the Experiment section that our method outperforms classic matrix factorization approach without contextual information.

## II. Related Work

Movie recommendation is a popular research topic and industrial product widely encountered in the industry. YouTube has published a series of video recommendation research papers, from its earliest shallow models [1] to the cutting-edge deep learning approaches [2]. In the larger picture, recommendation system has a broad spectrum of framework and algorithms, all of which could be tested for the movie recommendation scenario. One of the popular recommendation frameworks is matrix factorization. Matrix factorization algorithms include SVDFeature [4], Alternating Least Squares [5], MatRec [6], and Zipf Matrix Factorization [7], among many proposed innovativeinventions. In 2021, Wang [3] proposed a new framework to replace the space-consuming tensor factorization paradigm called MatMat, which is the building block of our contextual movie recommendation system. The algorithm can be executed by stochastic gradient descent. It is also highly efficient in time complexity.

Context-aware recommendation is a subfield of recommender system that is an actively researched hot topic. Techniques such as kernel method [8] and tensor factorization [9] have been proposed to enhance recommendation with contextual information. Research on context-aware recommendation is usually very difficult to carry out due to the data collection problem. Thanks to *University of Ljubljana,* we have access to the LDOS-CoMoDa dataset [10] that contains user rating information together with contextual information such as location and mood and weather.

### III. MATMAT FORMULATION

The fundamental idea behind matrix factorization is as follows: We decompose the user item rating matrix into the dot product of user feature vectors and item feature vectors as a dimensionality reduction framework to boost the accuracy of recommender system. The space complexity of the framework is reduced to O(m+n) instead of O(mn) by this approach, where m is the number of users and n is the number of items. The time complexity is minimized using stochastic algorithms such as Stochastic Gradient Descent. The formal classic matrix factorization problem formulation is shown below :

$$\text{RMSE} = \sum_{i=1}^{m}\sum_{j=1}^{n}\left(R_{ij} - u_i^T \cdot v_j\right)^2$$

where $R_{ij}$ is the user item rating value, $u_i$ is the user feature vector, and $v_j$ is the item feature vector. After computing for the optimal values of $u_i$ and $v_j$, the unknown user item rating is constructed using the dot product of the two parameter vectors.

There have been a great number of variants of matrix factorization framework, ranging from SVDFeature, Alternating Least Squares, to MatRec and Zipf Matrix Factorization. In 2021, Wang [7] proposed a new framework for matrix factorization based multitask learning and contextual recommendation framework. The author replaces the scalar values to be fit in the user rating matrix by matrices, namely:

$$R = \begin{bmatrix} \begin{bmatrix} r_{1,1} & \cdots & \cdots \\ \cdots & \cdots & \cdots \\ \cdots & \cdots & r_{1,1} \end{bmatrix} & \cdots & \begin{bmatrix} r_{1,n} & \cdots & \cdots \\ \cdots & \cdots & \cdots \\ \cdots & \cdots & r_{1,n} \end{bmatrix} \\ \cdots & \cdots & \cdots \\ \begin{bmatrix} r_{n,1} & \cdots & \cdots \\ \cdots & \cdots & \cdots \\ \cdots & \cdots & r_{n,1} \end{bmatrix} & \cdots & \begin{bmatrix} r_{n,n} & \cdots & \cdots \\ \cdots & \cdots & \cdots \\ \cdots & \cdots & r_{n,n} \end{bmatrix} \end{bmatrix}$$

where each element of R is a matrix whose diagonal values are replicates of the user item rating matrx elements of the classic matrix factorization framework at the same positional index. The off-diagonal values of each element are contextual information variables for recommendation.

By this formulation, the matrix factorization problem becomes a matrix fitting problem where the submatrices of R are fitted by user feature matrix and item feature matrix. The framework could be solved using standard optimization techniques such as Stochastic Gradient Descent and Adagrad. After the computation of parameters, the unknown matrix elements including user item rating values and contextual information are predicted by matrix multiplications of the user feature matrix and item feature matrix.

The major intuition behind MatMat is to find a simple and efficient way to replace more complex models such as tensor factorization. Tensor factorization is notorious for its high consumption of data storage. Even the most advanced commercial computing facilities could not meet the requirement for a million user product. By resorting to modification of matrix factorization framework, MatMat is able to reduce the space complexity of algorithmic parameters to O(m+n), where m is the number of users and n is the number of items. In addition, to store the input dataset, MatMat consumes only $O(k^2N)$, where k is the size of each square matrix element, and N is the number of user item rating values. The space complexity required is much less than tensor factorization.

Since in the MatMat framework, the classic user item rating matrix has been converted into a matrix of matrices, each element of the matrix can incorporate multiple contextual information factors. The interaction among variables for matrix fitting is nonlinear and can reflect highly complicated relations since the dot product of user and item feature vectors now becomes matrix product of two matrices. In the prediction stage of the algorithm, not only the user item rating can be predicted, the contextual information can be estimated as well.

In the MatMat paper, Wang [7] proposes a real world example that incorporates rank information into the matrix to be fitted. The researcher conducts experiments on MovieLens small datasets and observes that by MAE score and fairness metrics, the particular example of MatMat is competitive with the classic matrix factorization model. In addition, the efficiency of MatMat is also guaranteed since fast optimization techniques such as Stochastic Gradient Descent could be used to solve for the optimal parameters.

### IV. MOVIEMAT FORMULATION

We propose a context-aware movie recommender system in our paper because it's very difficult to find a real world example to illustrate the usefulness of MatMat framework. In addition, movie recommendation is one of the most promising scenarios in which context-aware recommendation can be applied. In this paper, we propose 2 context-aware movie recommendation algorithms based on the idea. One is called MovieMat, the other is called MovieMat+, which uses more contextual information.

To fit in the framework of MatMat, we define the submatrices of R as follows:

$$R_{i,j} = \begin{bmatrix} \dfrac{r_{i,j}}{max(r)} & \dfrac{location}{max(location)} \\ \dfrac{mood}{max(mood)} & \dfrac{r_{i,j}}{max(r)} \end{bmatrix}$$

The location and mood are scalarized values related to user-item contextual environment. Location represents the type of physical environment when watching the movie, while mood represents the mood of the audience.

We use Stochastic Gradient Descent algorithm to solve for the optimal values of the user feature matrix and item feature matrix. The loss function to be optimized is as follows:

$$\text{RMSE} = \sum_{i=1}^{m}\sum_{j=1}^{n} ||U_i^T \cdot V_j - R_{i,j}||^2$$

where $R_{ij}$ is the target submatrix element, $U_i$ is the user feature matrix, and $V_j$ is the item feature matrix. After computing for the optimal values of $U_i$ and $V_j$, the unknown user item rating is constructed using the matrix product of the two parameter matrices.

The remaining challenge is how to collect contextual data such as location and mood. Actually unlike other public dataset, it is extremely difficult to acquire test dataset for context aware recommender system. Luckily we are able to gain access to LDOS-CoMoDa dataset [10] provided by *University of Ljubljana*, where there is rich information of event contexts. We illustrate the superiority of our approach compared with non-contextual matrix factorization in the next section.

For MovieMat+, we consider more contextual information. Namely, we incorporate 7 contextual information types into the submatrices of R as follows:

$$R_{i,j} = \begin{bmatrix} \frac{r_{i,j}}{max(r)} & \frac{daytype}{max(daytype)} & \frac{season}{max(season)} \\ \frac{weather}{max(weather)} & \frac{r_{i,j}}{max(r)} & \frac{location}{max(location)} \\ \frac{emotion}{max(emotion)} & \frac{mood}{max(mood)} & \frac{r_{i,j}}{max(r)} \end{bmatrix}$$

Detailed explanation of meanings of the side information can be found in the description file of the LDOS-CoMoDa test dataset. They are mainly the location, weather, audience's emotions, audience's mood, weather and daytype. All contextual information are scalarised categorical values that can be normalized and analytically computed with other variables.

MovieMat and MovieMat+ are 2 practical examples of MatMat applications. We could naturally add more or different contextual information variables in the off-diagonal positions of the matrix R if more or different information becomes available. Context-aware movie recommendation can be used in movie theatres to enhance user experiences of users. If the theatre has smart sensors installed, it is very easy to have access to th audience' contextual information. A big data and AI empowered theatre industry could become the next generation theatre standards.

## V. EXPERIMENT

We test MovieMat and MovieMat+ on LDOS-CoMoDa dataset that contains 121 users and 1232 movies. For the first experiment, we use contextual information location (scalar values from 1 to 3) and user mood (scalar values from 1 to 3) to test MovieMat against the classic matrix factorization algorithm. The evaluation metrics are MAE (Fig.1) for technical accuracy and Degree of Matthew Effect [7] for fairness evaluation (Fig.2). Both algorithms tested in this section are resolved by Stochastic Gradient Descent. The results of the 2 algorithms are compared by doing grid search on gradient learning step values:

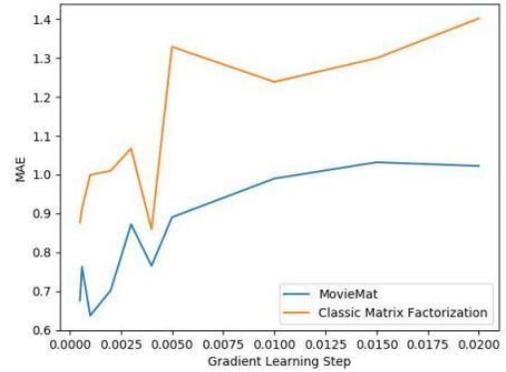

Fig. 1. Comparison between MovieMat and the Classic Matrix Factorization Algorithm on the MAE score

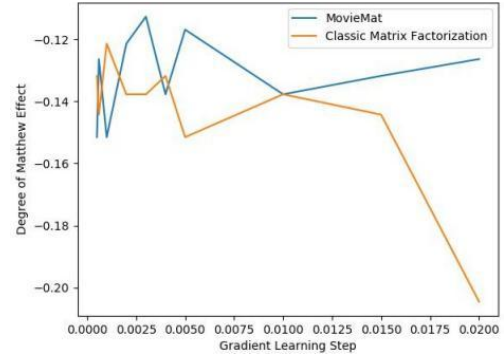

Fig. 2. Comparison between MovieMat and the Classic Matrix Factorization Algorithm on the Degree of Matthew Effect

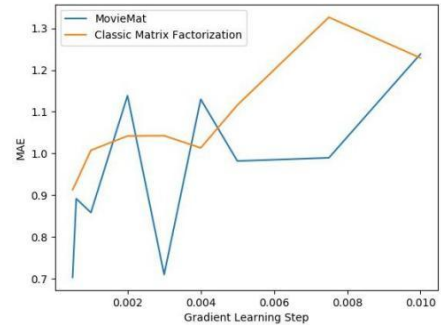

Fig. 3. Comparison between MovieMat and the Classic Matrix Factorization Algorithm on the MAE score

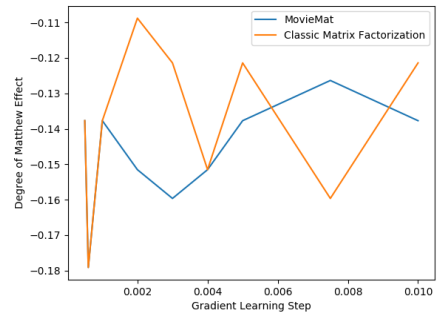

Fig. 4. Comparison between MovieMat+ and the Classic Matrix Factorization Algorithm on the Degree of Matthew Effect metric

The experiments show that the MAE of MovieMat can achieve a score lower than 0.7 at its best, which is superior to the performance of the classic matrix factorization by a large margin. When evaluated in fairness metric, MovieMat also outperforms the classic matrix factorization paradigm. MovieMat+ (Fig.4) also outperforms the classic matrix factorization paradigm, but is slightly inferior to MovieMat with the best MAE score slightly above 0.7.

MovieMat is very fast and computationally feasible in modern computing environments. We conducted our experiments on a Lenovo laptop with 1.8GHz AMD A6-6310 quad-core CPU and 12GB memory. The experiments are accomplished on the scale of seconds.

## VI. Discussion

By incorporating contextual information into movie recommeder systems, we observe improvement in both accuracy and fairness metrics. Contextual information by MovieMat and MovieMat+ can be viewed as a particular case of multi-task learning. It is good to know that multi-task learning does not necessarily produce trade-off between accuracy and multi-task goals.

A strikingly different merit of MovieMat and MovieMat+ than tensor factorization and other context-aware models is that it allows more complicated relation in interaction between variables. We believe the complexity of interaction can lead to better performance compared with other models in general.

Recently researchers [12] pointed out that there exists a unified foundation underlying matrix factorization and linear models. We would like to explore relations between linear models and MatMat-based recommender systems as well. We hope we could find a unfied theoretical relation underlying MatMat and other models as well.

We notice that in our experiments, incorporating more information does not necessarily lead to better results. We would like to investigate into the relation between contextual information type / quantity and accuracy / fairness. We want to anwser questions such as "Is there a singular point of the number of contextual variables beyond which the increase of accuracy levels off ? "

## VII. Conclusion

In this paper, we propose a new movie recommender system called MovieMat that takes advantage of MatMat framework and contextual information of location and audience mood to recommend interesting movies to the audience. The method outperforms the classic matrix factorization without contextual information by large margins and is computationally efficient in commercial computing environments.

One concern about MovieMat (MovieMat+) and MatMat in general is the cold-start problem. Wang [11] proposes a matrix factorization approach that takes no input data and generates competitive cold-start suggestions. The technique can be applied in movie recommendation settings as well. But a natural question is what can be achieved if we have partial information of the MatMat score matrix? Is there a different cold start algorithm for MatMat that is superior to ZeroMat?

Another issue in question is the fairness issue. Is there a way to enhance the fairness performance of MatMat based algorithms similar to Zipf Matrix Factorization [7] and other algorithms? What is the upper limit of performance of MatMat based algorithms?

In future work, we would like to explore further the potentiality of the MatMat framework applied in the scenario of contextual recommendation, to answer the previous questions. Moreover, we would like to explore the optimization techniques and performance of MovieMat. We wish our work could lead to future technical revolution in the theatre industry.